\documentclass[twocolumn,showpacs,preprintnumbers,amsmath,amssymb,prb]{revtex4}
\usepackage{graphics}
\usepackage{dcolumn}
\usepackage{color}
\usepackage{bm}
\begin{document}

\title{Structure, electronic properties and magnetic transition in manganese clusters}
\author{Mukul Kabir}
\affiliation{Unit for Nanoscience and Technology, S.N. Bose National Center for Basic Sciences, JD Block, Sector III, Salt Lake 
City, Kolkata 700 098, India}
\author{D. G. Kanhere}
\affiliation{Department of Physics and Center for Modeling and Simulation, \\ 
University of Pune, Pune - 411 007, India}
\author{Abhijit Mookerjee}
\affiliation{Unit for Nanoscience and Technology, S.N. Bose National Center for Basic Sciences, JD Block, Sector III, Salt Lake 
City, Kolkata 700 098, India}

\date{\today}

\begin{abstract}
We systematically investigate the structural, electronic and magnetic 
properties of Mn$_n$ clusters ($n =$ 2$-$20) within the {\it ab-initio}
pseudopotential plane wave method using generalized gradient approximation
for the exchange-correlation energy. A new kind of icosahedral structural
growth has been predicted in the intermediate size range. Calculated 
magnetic moments show an excellent agreement with the Stern-Gerlach 
experiment. A transition from ferromagnetic to ferrimagnetic Mn$-$Mn coupling
takes place at $n=$ 5 and the ferrimagnetic states continue to be the ground states for the entire size range. Possible
presence of multiple isomers in the experimental beam has been argued.
No signature of non-metal to metal transition is observed in this size
range and the coordination dependence of $d-$electron localization is 
discussed. 
\end{abstract}
\maketitle

\section{\label{sec:intro}Introduction}
The search for magnetic behavior in the transition metal
clusters is motivated largely by the desire to understand how
magnetic properties change in  the reduced dimension. This is a question of
considerable technological importance. Several unexpected magnetic orderings 
have already been reported in the reduced dimension. This ranges from the prediction
of net magnetic moment in  clusters of nonmagnetic (Rh, Ref. 1) or
antiferromagnetic (Cr, Ref. 2 and Mn, Refs. 3 and 4) bulk materials 
to the enhancement in magnetic moment in  clusters of ferromagnetic 
metals (Fe, Ref. 5 and Co, Ref. 6).

Manganese clusters are particularly interesting among all 3$d$
transition metal elements due to the 4$s^2$, 3$d^5$ electronic
configuration in Mn atoms. Because of the filled 4$s$ and half-filled 3$d$ shells
and the large energy gap $\sim$ 8 eV between these levels and as well
as due to the high 4$s^2$3$d^5$ $\rightarrow$ 4$s^1$3$d^6$
promotion energy of 2.14 eV\cite{moore},  Mn atoms do not bind strongly. As a result, Mn$_2$ is
a weakly bound van der Waals dimer with reported bond dissociation energy ranging
from 0.1 $\pm$ 0.1 to 0.56 $\pm$ 0.26 eV depending upon the different method of 
analysis\cite{morse,gingerich,kant,teraski}. This weak Mn$-$ Mn bonding has been demonstrated 
through the photodissociation experiments for Mn$^+_n$ ($n \le 7$) cluster cations\cite{teraski,tono}.
Consequently, the bulk $\alpha$-Mn, which has a very complex lattice structure with 58 atoms 
in the unit cell, has the lowest binding energy among all the 3$d$ transition metal elements.

Magnetic properties of manganese clusters are rather unusual. According to 
Hund's rule, the half-filled
localized 3$d$ electrons give rise to large atomic magnetic moment of 5
$\mu_B$. An early electron spin resonance
(ESR) study suggested a magnetic moment of 5 $\mu_B$/atom for
very small Mn clusters\cite{zee}.  However, Stern-Gerlach (SG) molecular
beam experiments on Mn$_{5-99}$ clusters  by Knickelbein recently
revealed the net magnetic moments ranging from 0.4 to 1.7 $\mu_B$/atom\cite{mark1, mark2}. This 
differs both from the ferromagnetic (FM) small clusters and from the
antiferromagnetic (AFM) bulk $\alpha$-Mn. This experimental results can be
explained either way that the individual atomic moments are small and
ordered ferromagnetically or the individual atomic moments remain
large but their orientation flips from site to site i.e., they are
coupled ferrimagnetically. In the SG experiment, it is important
to note the relative decrease in the magnetic moment for Mn$_{13}$
and Mn$_{19}$, as well as the relatively very large experimental
uncertainty in the measured magnetic moment for Mn$_7$\cite{mark1, mark2}. In the present
work, we will show that the local minima for Mn$_{13}$ and Mn$_{19}$
arise due to their `closed' icosahedral structures, whereas, the large experimental
uncertainty ($\pm$ 58 \% of the measured value) in the magnetic moment of Mn$_7$ is
plausibly due to the production of different magnetic isomers (in addition with 
statistical fluctuation) in the subsequent measurements. 

Earlier all electron (AE) studies \cite{jena, nayak, pederson} found Mn$-$Mn
FM ordering for Mn$_n$ ($n > $ 4) clusters, which, in turn, is not
consistent with the SG experiment. 
However, Nayak {\it et al.} first predicted ferrimagnetic ground state for Mn$_{13}$ with
a total magnetic moment of 33 $\mu_B$\cite{nayak2}.
In consistent with the SG experiments, 
very recent AE studies by Parvanova {\it et al.}\cite{bobadova1,
bobadova2} ($n=$2-9, 12 and 13) and Jones {\it et al.}  \cite{jones}
($n=$ 5 and 6) reported ferrimagnetic ordering in Mn$_n$ clusters.
 Briere {\it et al.}  \cite{briere}
used ultra-soft pseudopotentials (US-PP) to study the intermediate
size Mn$_n$ clusters ($n$ = 13, 15, 19 and 23) and found icosahedral
structural growth with an exception for Mn$_{15}$. 
However, their predicted magnetic moments differ widely from
the experimental values. This might be attributed to the reason that
the US-PP may not be appropriate in describing  the
transition metals with large magnetic moments. This will be discussed briefly
later in the section \ref{sec:methodology}. 
Our main motivation of this work is particularly driven by the SG experiments\cite{mark1, mark2}.
Here we shall investigate $-$ (i) The structural and magnetic
evolution of  Mn$_n$ clusters, $n$=2-20.  (ii) The sudden drop in
the magnetic moment at $n$=13 and 19 and the very large experimental
uncertainty in the measured magnetic moment for Mn$_7$, and (iii) The possible presence 
of isomers with different magnetic
structures in the SG experimental molecular beam.

It has also been found by Parks {\it et al.} that the Mn$_n$ clusters
show a downward discontinuity in their reaction rate with molecular hydrogen at
$n =$ 16, and this was  attributed to non-metal to metal 
transition in Mn$_{16}$\cite{parks}. But if this is indeed true then there should be 
a downward decrease in the ionization potential too. However, no such 
abrupt decrease has been seen in the measured ionization potential\cite{koretsky}.
In the present paper, we calculate both the spin gaps to investigate 
this issue.

\section{\label{sec:methodology} Computational Details}
The calculations are performed using density functional theory (DFT), within the pseudopotential
plane wave method. We have used projector augmented wave (PAW) method
\cite{blochl, kresse} and Perdew-Bruke-Ernzerhof (PBE)
exchange-correlation functional \cite{perdew} for spin-polarized
generalized gradient correction (GGA) as implemented in the Vienna {\it ab-initio} Simulation
Package ({\small VASP})\cite{kresse2}.  The 3$d$ and 4$s$ electrons are treated as
valence electrons and the wave functions are expanded in the plane
wave basis set with the kinetic energy cut-off 337.3 eV. Reciprocal
space integrations are carried out at the $\Gamma$ point. Symmetry
unrestricted geometry and spin optimizations are performed using
conjugate gradient and quasi-Newtonian methods until all the force
components are less than a threshold value 0.005 eV/\r{A}. Simple
cubic supercells are used with the periodic boundary conditions,
where two neighboring clusters are kept separated by at least 12 \r{A}
vacuum space. For each size, several initial geometrical structures have been
considered. To get the ground state magnetic moment we have explicitly
considered {\it all possible} spin configurations for each geometrical
structure. For transition metals with large magnetic moments,
the PAW method seems to be  more appropriate (as good as the AE calculations) than
the US-PP approach\cite{kresse}.  The US-PP overestimates the
magnetization energies and this overestimation is even more large for
GGA calculations than local spin density approximation (LSDA).  This is due
to the fact that the GGA functionals are more sensitive to the shape
of the wave functions than the LSDA functionals. However, the difference
between these two methods, US-PP and PAW, are solely related to the
pseudization of the augmentation charges in the US-PP approach, which
can be removed by choosing very accurate pseudized augmentation
function, which is then computationally expensive.  For a better
description see the Ref. 25 by Kresse and Joubert.

The binding energy per atom is calculated as,
\begin{equation}
E_b(\mbox{Mn$_n$}) \ = \ \frac{1}{n}\left[ E(\mbox{Mn$_n$}) \ - \ n \ E(\mbox{Mn}) \right],
\end{equation}
$n$ being the size of the cluster. The local magnetic moment $\mathcal{M}$, at each site 
can be calculated as,
\begin{equation}
\mathcal{M} \ = \ \int_0^R \ \left[ \ \rho^{\uparrow}(\mathbf {r}) \ - \ \rho^{\downarrow}(\mathbf {r}) \right] \ d\mathbf {r},
\end{equation}
where $\rho^{\uparrow}(\mathbf {r})$ and $\rho^{\downarrow}(\mathbf
{r})$ are spin-up and spin-down charge densities, respectively and $R$
is the radius of the sphere centering the atom. For a particular
cluster, R is taken such that no two spheres overlap i.e., R is equal
to the half of the shortest bond length in that cluster.

\section{\label{sec:results}Results and discussions}
\subsection{\label{sec:Mn2-4}Small Ferromagnetic Clusters (Mn$_2$ - Mn$_4$)}

The Mn$_2$ dimer is the most controversial among all the sizes we
have studied. Experiments based on resonance Raman spectroscopy
\cite{bier} and ESR \cite{zee2} observed an AFM
ground state with a bond length 3.17 \r{A}. Experimentally the binding energy was
estimated to be 0.44 $\pm$ 0.30 eV/atom \cite{morse,gingerich,kant}. However,
previous AE-DFT calculations \cite{jena, pederson,
bobadova1, bobadova2}, predicted a FM state to be the
ground state with much smaller bond length ($\sim$ 2.60 \r{A}) than that of
the experimental value. In agreement with these calculations, our
present PAW pseudopotential calculations with PBE exchange-correlation
functionals, predict a FM ground state with total spin 10 $\mu_B$ and
with bond length 2.58 \r{A}. Very small binding energy, 0.52 eV/atom,
is essentially the characteristic of a van der Waals system. 
However, the bond dissociation energy increases by considerable amount due to
the reduction of one electron from the Mn$_2$ dimer, i.e. by creating a hole
in the 4$s$ level. As measured by the photodissociation experiment the 
bond dissociation (Mn$^+$ $\cdots$ Mn) energy increases to 1.39 eV for Mn$^+_2$
cation\cite{jarrold}. We find that the bond length of Mn$_2$ decreases 
monotonically as the net moment decreases. This is consistent with the 
physical picture that the reduction of the interatomic spacing leads to 
comparatively stronger overlap of the atomic orbitals which, in turn, reduces the
magnetic moment. However, we find Mn$_2$ with total moment 4 and
6 $\mu_B$ are not bound. An AF Mn$_2$ is 0.52 eV higher in energy with
bond length 2.61 \r{A}.  The present results, along with the previous
AE-DFT calculations \cite{jena, pederson,bobadova1, bobadova2}, do not
agree with the experimental results. 
This might be attributed to the fact that the density functional theory is not
adequate in treating van der Waals interaction properly.
However, there is no experimental
results available in the gas phase Mn$_2$ dimer and the ESR experiment was
done in a rare gas matrix. Therefore, it is possible that the Mn atoms
do interact with the matrix, which could stretch the bond length and
could lead to the observed AFM state.

\begin{figure}[!t]
{\rotatebox{0}{\resizebox{8cm}{6cm}{\includegraphics{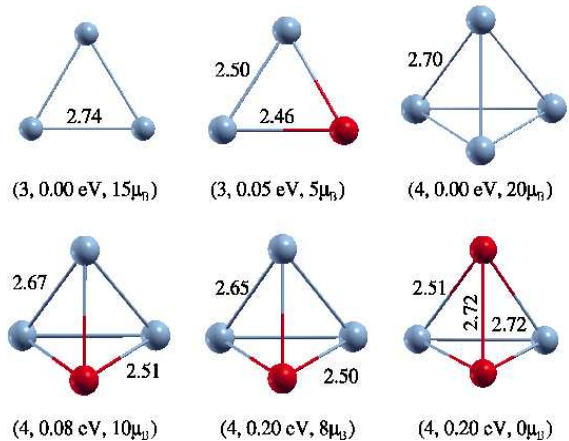}}}}
\caption{\label{fig:2to4} (Color online). Atomic spin ordering of the ground
state and low-lying isomers for Mn$_3$ and Mn$_4$
clusters. Numbers in the parenthesis represent number of atoms in the
cluster, relative energy to the ground state and total magnetic
moment, respectively. Bond lengths are given in \r{A}. Blue (Gray) color
represents up or positive and red (dark gray) represents down or negative 
magnetic moment. We will follow the same convention throughout.}
\end{figure}

The case of Mn$_3$ is extremely interesting as it could have either FM
or a frustrated AFM structure.  We have studied triangular and linear
structures. An equilateral triangular
FM state with total moment 15 $\mu_B$ is found to be the ground state
with bond lengths 2.74 \r{A} and binding energy 0.82 eV/atom. The
frustrated AFM state with total moment 5 $\mu_B$ is found to be nearly degenerate
with the FM ground state (lies only 0.05 eV
higher in energy). This has an isosceles triangular structure with one
long and two short bond lengths of 2.50 and 2.45 \r{A},
respectively (Fig.\ref{fig:2to4}). The resonance Raman spectra studies by Bier {\it et al.}
\cite{bier} suggest the ground state to be Jhan-Teller distorted
$D_{3h}$ structure with an odd integer magnetic moment.

\begin{figure}[!t]
{\rotatebox{0}{\resizebox{8cm}{17cm}{\includegraphics{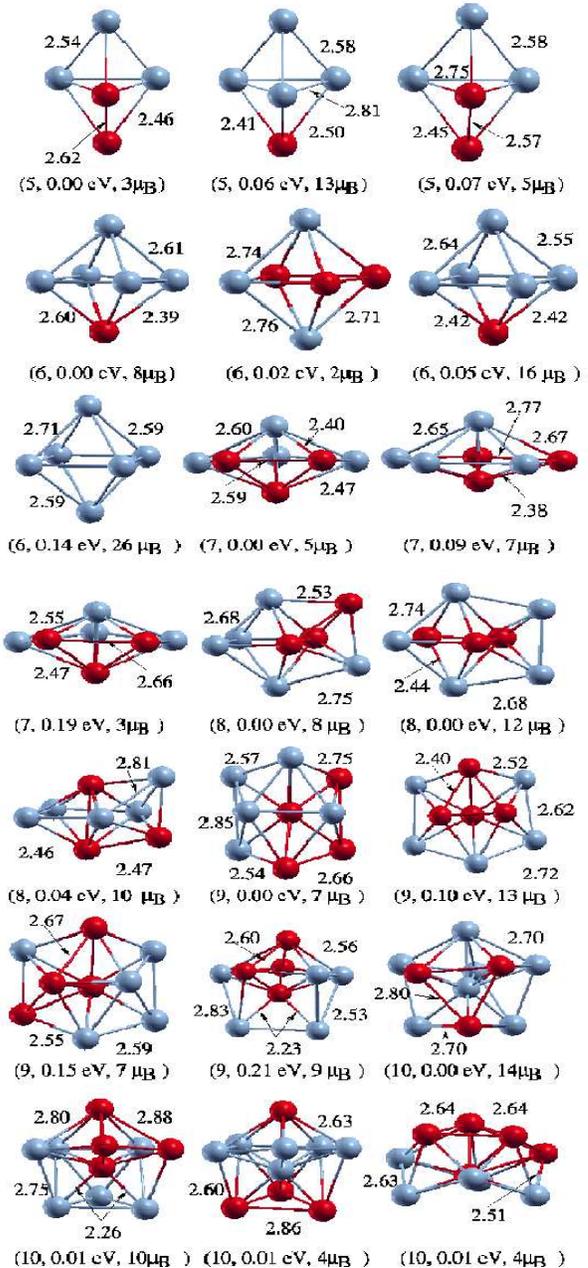}}}}
\caption{\label{fig:5to10} (Color online). Atomic spin ordering of the ground and 
isomeric geometries for $n =$ 5-10. Same ordering has been followed as in the 
Table \ref{tab:bemag}.}
\end{figure}

For the Mn$_4$ cluster, we examined three different conformations:
square, rhombus and tetrahedron. A perfect
tetrahedral structure with bond lengths 2.7 \r{A} and binding energy
1.18 eV/atom is the ground state, where Mn$-$Mn coupling is FM with
total moment 20 $\mu_B$ (Fig.\ref{fig:2to4}). Three isomers are found and all of them are
tetrahedral. A ferrimagnetic state with total moment 10 $\mu_B$ is only
0.08 eV higher in energy. Another ferrimagnetic state with total moment
8 $\mu_B$ is found to be degenerate with the AF state with no net
moment and they are 0.20 eV higher in energy. In all these optimal
structures the distances between two similar spins ($d_{\uparrow
\uparrow}$ or $d_{\downarrow \downarrow}$) are larger than those of
between two opposite spins ($d_{\uparrow \downarrow}$).  Our results
are consistent with the previous AE calculations\cite{jena, pederson,
bobadova1, bobadova2}.  Ludwig {\it et al.} \cite{ludwig} have studied
Mn$_4$ in solid silicon and observed a 21-line hyperfine pattern that
not only establishes the four atoms to be equivalent, but also the
total moment to be 20 $\mu_B$. However, the present results can not
directly be compared with this experiment because of possible Si-Mn
interaction and there is no available report of magnetic ordering for
Mn$_4$ in its gas phase.

\subsection{\label{sec:Mn5-10}Mn$_5$ - Mn$_{10}$}
As the number of atoms in the cluster ($n$) increases the determination of the 
structural and magnetic ground state becomes a  very delicate task as the number of 
local minima in the corresponding potential energy surface increases exponentially
with $n$. Therefore, more than one geometric and/or magnetic structures of comparable
stability are possible. In the Fig.\ref{fig:5to10} we depict the atomic and magnetic
structures for the ground state as well as for the closely lying isomers for the size range
$n =$ 5-10.
As it is mentioned earlier, to hit the ground state more reliably, 
we have studied {\it all possible} spin
multiplicities for several geometric structures for a particular
cluster size $n$.  
Calculated binding energies, relative energies, magnetic moments and two spin gaps
are given in the Table \ref{tab:bemag} for the entire size range $n =$ 2-20.

For the Mn$_5$ cluster, a square pyramid and a
triangular bi-pyramid (TBP) were studied. Transition in the magnetic
order, from FM to ferrimagnetic, is found. A
ferrimagnetic TBP is found to be the ground state with total spin 3
$\mu_B$. The next two isomers are also ferrimagnetic in nature with
total spins 13 $\mu_B$ and 5 $\mu_B$. Both of these structures also have TBP
structure and lie 0.06 and 0.07 eV, respectively, higher in energy. The next lowest
energy arrangement is FM and also has a TBP structure with total spin 23
$\mu_B$ and lies 0.19 eV higher in energy.  Our results are in
agreement with the very recent AE calculations\cite{bobadova1,
bobadova2, jones}.  However, previously the FM ground state was predicted by both
Nayak {\it et al.}\cite{jena, nayak} and Pederson {\it et al.} \cite{pederson}. 
In the recent SG experiment \cite{mark2}, magnetic moment
was found to be 0.79$\pm$0.25 $\mu_B$/atom, which is very close to our
predicted value 0.60 $\mu_B$/atom for the ground state.

We have investigated both the octahedral and the capped trigonal
bi-pyramid for Mn$_6$ cluster. A ferrimagnetic octahedral structure with total spin 8
$\mu_B$ is found to be the ground state with binding energy 1.57
eV/atom. Another octahedral ferrimagnetic isomer with total moment 2
$\mu_B$ is nearly degenerate (0.02 eV higher in energy).  The next
isomer is also a ferrimagnetic octahedra, which possess a total moment of 
16 $\mu_B$ and lies 0.05 eV higher. The next favorable
isomer is FM and has a total moment 26 $\mu_B$ and is 0.14 eV higher
than the ground state.  In an earlier calculation, Pederson {\it et
al.} \cite{pederson} predicted a FM octahedral structure with
moment 4.33 $\mu_B$/atom to be the ground state. However, in agreement
with the recent AE-DFT calculations \cite{bobadova1, bobadova2, jones}, 
present calculation predicts the same ground state and
isomers.  Experimentally measured magnetic moment 0.55
$\pm$ 0.10 $\mu_B$/atom\cite{mark2} lies between that of our predicted ground
state, 1.33 $\mu_B$/atom and the first isomer, 0.33 $\mu_B$/atom,
which are almost degenerate. It is possible that in the SG
experimental beam, multiple isomers were produced such that the
measured value is almost an average of the ground state and the first
isomer.

We have considered pentagonal bi-pyramid (PBP), capped octahedron and
bi-capped trigonal pyramid as the possible candidates for the ground
state of Mn$_7$. The most stable configuration is a PBP structure with
ferrimagnetic spin ordering, which has a total moment 5 $\mu_B$. The next two
closest isomers also have ferrimagnetic arrangements with 7
$\mu_B$ and 3 $\mu_B$ total moments and they lie 0.09 eV and 0.20 eV higher
than the ground state, respectively. Our ground state magnetic moment agrees with
the earlier calculations \cite{bobadova1, bobadova2, jena2}, though we
predict isomers with different spin arrangements. However, Pederson
{\it et al.} predicted a FM ground state \cite{pederson}. Present
ground state magnetic moment per atom exactly matches with the
experimental value, 0.72 $\pm$ 0.42 $\mu_B$/atom \cite{mark2}. 
We would like to note the rather large uncertainty here. 
We argue that the
plausible presence of these isomers, with total moments 7 $\mu_B$ and 3 $\mu_B$
along with the ground state (5 $\mu_B$), in the SG beam might lead to this
high uncertainty in the measured value.

Motivated by our earlier study on Cu$_8$\cite{kabir3, kabir4}, we investigated three
different geometries for Mn$_8$, viz, capped pentagonal bi-pyramid
(CPBP), bi-capped octahedron (BCO) and tri-capped trigonal bi-pyramid
(TCTBP). The BCO structure with total moments 8 $\mu_B$ and 12 $\mu_B$
are found to be degenerate ground state. Another BCO structure with
total moment 10 $\mu_B$ lies only 0.03 eV higher in energy. The SG
cluster beam experiment has reported a magnetic moment of 
1.04 $\pm$ 0.14 $\mu_B$/atom\cite{mark2}, which is nearly an average of our
predicted values. Therefore, our present DFT study together with 
the experiment in turn indicate the possible presence of these three
isomers in the experimental beam with almost equal statistical
weight. The optimal CPBP and TCTBP structures have total moments 14
$\mu_B$ and 12 $\mu_B$, respectively and they lie 0.31 and 0.4 eV higher than the ground
state. Parvanova {\it et. al.} \cite{bobadova2} found CPBP structure
to be the most stable, however, their predicted magnetic moment is
very small, 4 $\mu_B$, compared to both of our value and the SG 
experiment. Pederson {\it et. al.} \cite{pederson} predicted a FM BCO
structure with moment 32 $\mu_B$ as the ground state. The optimal
FM structure for all the three geometrical structures have total
moment 32 $\mu_B$ and lie 1.01, 0.63 and 1.17 eV higher in energy
compare to their respective optimal ferrimagnetic structure,
respectively for BCO, CPBP and TCTBP structures.

For the Mn$_9$ cluster, as initial configuration we took three stable
isomers found for Cu$_9$\cite{kabir3, kabir4} and a capped and a centered antiprism 
structure.
The optimal
structure is a centered antiprism structure with total moment 7
$\mu_B$, which is in very good agreement with the experimental value
1.01 $\pm$ 0.10 $\mu_B$/atom \cite{mark2}. 
The local magnetic moment $\mathcal{M}$ (as calculated form the Eq. 2) shows
strong environment dependency due to the anisotropy in bonding. 
The $\mathcal{M}$ of the highly coordinated central atom is very small, $-$0.22 $\mu_B$,
whereas those of the surface atoms are quite high and lie between 3.45
and 3.75 $\mu_B$.  Parvanova {\it et al.} \cite{bobadova2} have found
a similar structure but with different spin configuration with total
moment 9 $\mu_B$ to be the optimal structure. 
The next two isomers have the same geometry and have 13 and 7 $\mu_B$ 
total magnetic moment (Fig.\ref{fig:5to10} and Table \ref{tab:bemag}). The next isomer is a
 bi-capped pentagonal bi-pyramid, which 
lies 0.21 eV higher with a total moment of 9 $\mu_B$.
The optimal capped
antiprism structure lies 0.23 eV higher and has a total moment of 7
$\mu_B$. Note that, all these structures have 5 spin-up ($N_{\uparrow}$) atoms and 4 spin-down 
($N_{\downarrow}$) atoms.

Different tricapped pentagonal bi-pyramidal structures along with
different tetra capped octahedral structures were tried as initial
structures for Mn$_{10}$. Four isomers exist with almost the same
energy.  They lie within $\sim$ 0.01 eV energy (see Table \ref{tab:bemag} 
and Fig. \ref{fig:intermediate}). All of these have a pentagonal 
ring and could be derived by
removing 3 atoms from a 13-atom icosahedra. Ground state has a total
magnetic moment 14 $\mu_B$, which is very close to the SG experimental
value, 1.34 $\pm$ 0.09 $\mu_B$/atom\cite{mark2}.

\begin{table*}[!t] 
\caption{\label{tab:bemag}Binding energy, relative energy to the GS ($\triangle E = E - E_{GS}$), magnetic moment (with a 
comparison to the SG experiment\cite{mark1,mark2}) and different spin gaps, $\triangle_1$ and $\triangle_2$, for Mn$_n$ ($n=$ 2-20) clusters.}
{\begin{tabular}{cccccccccccccccc} 
\hline
\hline
Cluster &  $E_b$       & $\triangle E$ & \multicolumn{2}{c} {Magnetic Moment} & \multicolumn{2}{c} {Spin Gaps} & & 
Cluster &  $E_b$       & $\triangle E$ & \multicolumn{2}{c} {Magnetic Moment} & \multicolumn{2}{c} {Spin Gaps} & \\
        & (eV/atom) &   (eV)        & \multicolumn{2}{c} {($\mu_B$/atom)}  & \multicolumn{2}{c} {(eV)} & &
        & (eV/atom) &   (eV)        & \multicolumn{2}{c} {($\mu_B$/atom)}  & \multicolumn{2}{c} {(eV)} &       \\
        &           &               & Theory   &  SG Exp.\cite{mark1,mark2}   & $\delta_1$ & $\delta_2$ & &
        &           &               & Theory   &  SG Exp.\cite{mark1,mark2}   & $\delta_1$ & $\delta_2$ &  \\
\hline
Mn$_2$     &     0.52  &     0.00  &     5.00     &   $-$             &    0.95     & 1.31     &  &
Mn$_{12}$  &     2.08  &     0.00  &     1.33     & 1.72 $\pm$ 0.04   &    0.48     & 0.26    \\
           &     0.26  &     0.52  &     0.00     &                   &    0.47     & 0.47     &  &
           &     2.08  &     0.05  &     0.33     &                   &    0.40     & 0.30    \\
Mn$_3$     &     0.82  &     0.00  &     5.00     &   $-$             &    0.73     & 1.27     &  &
           &     2.07  &     0.11  &     1.50     &                   &    0.05     & 0.45    \\
           &     0.81  &     0.05  &     1.67     &                   &    0.63     & 0.58     &  &
Mn$_{13}$  &     2.17  &     0.00  &     0.23     & 0.54 $\pm$ 0.06   &    0.34     & 0.38    \\
Mn$_4$     &     1.18  &     0.00  &     5.00     &   $-$             &    0.66     & 2.35     &  &
           &     2.16  &     0.08  &     0.54     &                   &    0.36     & 0.20    \\
           &     1.16  &     0.08  &     2.50     &                   &    0.45     & 0.85     &  &
Mn$_{14}$  &     2.17  &     0.00  &     1.29     & 1.48 $\pm$ 0.03   &    0.23     & 0.24    \\
           &     1.13  &     0.20  &     0.00     &                   &    0.41     & 0.41     &  &
           &     2.17  &     0.02  &     1.43     &                   &    0.24     & 0.31    \\
           &     1.13  &     0.20  &     2.00     &                   &    1.12     & 0.21     &  &
           &     2.17  &     0.05  &     1.57     &                   &    0.25     & 0.32    \\
Mn$_5$     &     1.41  &     0.00  &     0.60     & 0.79 $\pm$ 0.25   &    1.03     & 0.30     &  &
Mn$_{15}$  &     2.23  &     0.00  &     0.87     & 1.66 $\pm$ 0.02   &    0.36     & 0.27    \\
           &     1.40  &     0.06  &     2.60     &                   &    0.97     & 0.37     &  &
           &     2.23  &     0.03  &     0.33     &                   &    0.16     & 0.29    \\
           &     1.40  &     0.07  &     1.00     &                   &    0.16     & 0.65     &  &
           &     2.23  &     0.06  &     0.47     &                   &    0.20     & 0.23    \\
           &     1.37  &     0.19  &     4.60     &                   &    0.55     & 0.77     &  &
           &     2.23  &     0.06  &     0.87     &                   &    0.27     & 0.36    \\
Mn$_6$     &     1.57  &     0.00  &     1.33     & 0.55 $\pm$ 0.10   &    0.48     & 0.35     &  &
           &     2.23  &     0.06  &     1.00     &                   &    0.25     & 0.45    \\
           &     1.56  &     0.02  &     0.33     &                   &    0.40     & 0.31     &  &
           &     2.21  &     0.28  &     0.47     &                   &    0.39     & 0.35    \\
           &     1.56  &     0.05  &     2.67     &                   &    0.86     & 0.32     &  &
Mn$_{16}$  &     2.27  &     0.00  &     1.25     & 1.58 $\pm$ 0.02   &    0.33     & 0.22    \\
           &     1.54  &     0.14  &     4.33     &                   &    0.98     & 1.16     &  &
           &     2.27  &     0.02  &     1.38     &                   &    0.19     & 0.52    \\
Mn$_7$     &     1.73  &     0.00  &     0.71     & 0.72 $\pm$ 0.42   &    0.45     & 0.65     &  &
           &     2.27  &     0.06  &     0.63     &                   &    0.28     & 0.35    \\
           &     1.71  &     0.09  &     1.00     &                   &    0.56     & 0.23     &  &
           &     2.27  &     0.10  &     0.50     &                   &    0.30     & 0.20    \\
           &     1.70  &     0.19  &     0.43     &                   &    0.51     & 0.13     &  &
Mn$_{17}$  &     2.33  &     0.00  &     1.59     & 1.44 $\pm$ 0.02   &    0.25     & 0.37    \\
Mn$_8$     &     1.77  &     0.00  &     1.00     & 1.04 $\pm$ 0.14   &    0.61     & 0.20     &  &
           &     2.32  &     0.08  &     1.47     &                   &    0.25     & 0.09    \\
           &     1.77  &     0.00  &     1.50     &                   &    0.40     & 0.41     &  &
           &     2.32  &     0.09  &     1.71     &                   &    0.14     & 0.70    \\
           &     1.77  &     0.04  &     1.25     &                   &    0.35     & 0.25     &  &
Mn$_{18}$  &     2.35  &     0.00  &     1.67     & 1.20 $\pm$ 0.02   &    0.36     & 0.30    \\
Mn$_9$     &     1.87  &     0.00  &     0.78     & 1.01 $\pm$ 0.10   &    0.49     & 0.36     &  &
           &     2.35  &     0.02  &     1.56     &                   &    0.34     & 0.33    \\
           &     1.86  &     0.10  &     1.44     &                   &    0.24     & 0.60     &  &
           &     2.35  &     0.02  &     1.44     &                   &    0.35     & 0.25    \\
           &     1.85  &     0.15  &     0.78     &                   &    0.30     & 0.34     &  &
           &     2.35  &     0.06  &     1.78     &                   &    0.18     & 0.55    \\
           &     1.84  &     0.21  &     1.00     &                   &    0.24     & 0.36     &  &
Mn$_{19}$  &     2.37  &     0.00  &     1.10     & 0.41 $\pm$ 0.04   &    0.19     & 0.22    \\
Mn$_{10}$  &     1.94  &     0.00  &     1.40     & 1.34 $\pm$ 0.09   &    0.27     & 0.44     &  &
           &     2.37  &     0.01  &     1.00     &                   &    0.24     & 0.16    \\
           &     1.94  &     0.01  &     1.00     &                   &    0.69     & 0.13     &  &
           &     2.37  &     0.08  &     0.47     &                   &    0.22     & 0.15    \\
           &     1.94  &     0.01  &     0.40     &                   &    0.36     & 0.41     &  &
Mn$_{20}$  &     2.37  &     0.00  &     1.40     & 0.93 $\pm$ 0.03   &    0.39     & 0.19    \\
           &     1.94  &     0.01  &     0.40     &                   &    0.37     & 0.20     &  &
           &     2.37  &     0.00  &     1.50     &                   &    0.21     & 0.20    \\
Mn$_{11}$  &     1.99  &     0.00  &     0.82     & 0.86 $\pm$ 0.07   &    0.26     & 0.29     &  &
           &     2.37  &     0.05  &     1.60     &                   &    0.12     & 0.35    \\
           &     1.98  &     0.11  &     0.46     &                   &    0.34     & 0.20     &  &
           &     2.37  &     0.07  &     0.80     &                   &    0.30     & 0.21    \\
           &     1.98  &     0.15  &     0.64     &                   &    0.10     & 0.45     &  &
           &           &           &              &                   &             &         \\
\hline
\hline
\end{tabular} }
\end{table*}

\subsection{\label{Mn10-20}Intermediate size clusters: Mn$_{11}$ - Mn$_{20}$}

\begin{figure*}[!t]
{\rotatebox{270}{\resizebox{9cm}{16cm}{\includegraphics{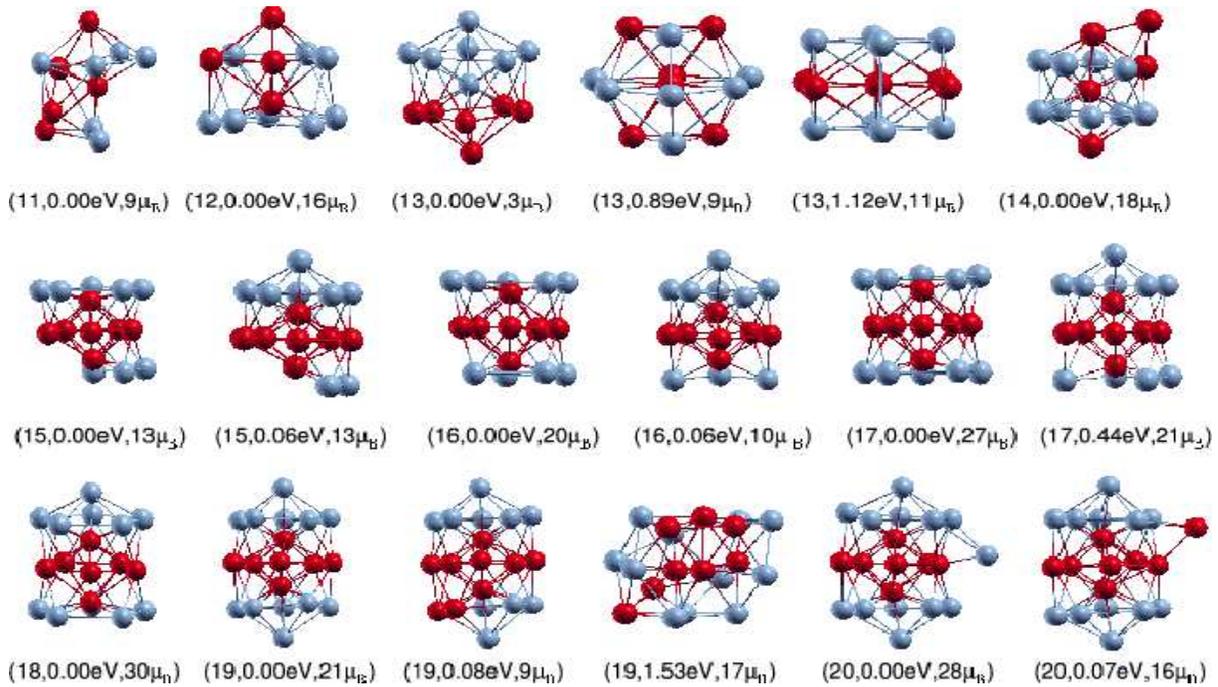}}}}
\caption{\label{fig:intermediate}(Color online) The ground state and a few higher energy structures for the size 
range $n =$ 11 $-$ 20. Note the grouping of the same kind of spins.}
\end{figure*}

\begin{figure}[b]
{\rotatebox{270}{\resizebox{6cm}{8.5cm}{\includegraphics{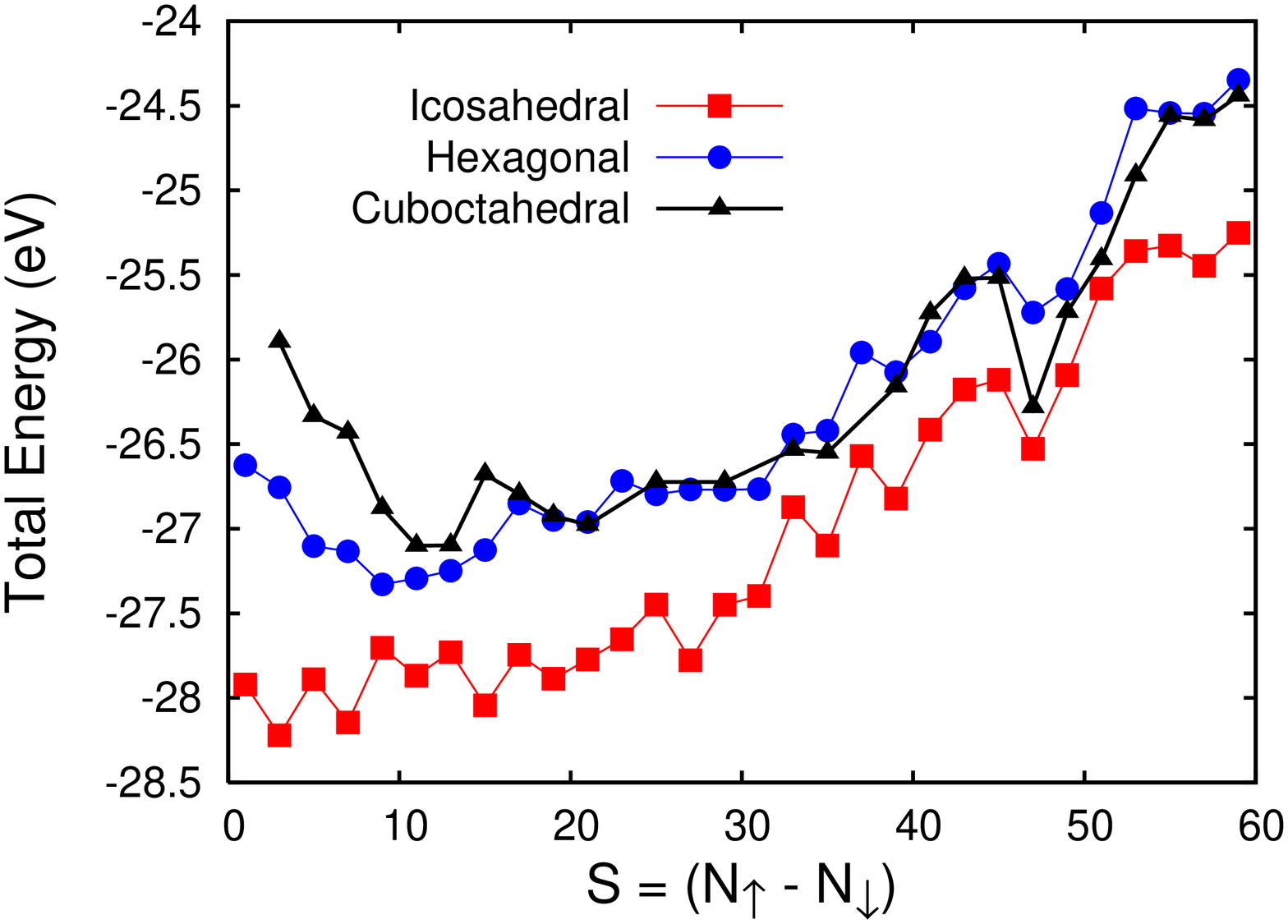}}}}
\caption{\label{fig:mn13} (Color online) Plot of the total energy as a function of total magnetic moment 
$S (= N_{\uparrow} - N_{\downarrow})$ for icosahedral, hexagonal closed pack and cuboctahedral
conformations for Mn$_{13}$ cluster.}
\end{figure}

All the intermediate sized clusters with $n =$ 11-20 are found to adopt  an icosahedral
growth pattern. The ground state structures and the few isomers along with their corresponding 
spin arrangements are
shown in the Fig.\ref{fig:intermediate}. An icosahedral structure without one apex atom is found
to be the ground state for the Mn$_{12}$ cluster. This structure has $N_{\uparrow}$ = 8
and $N_{\downarrow}$ = 4 spin configuration with a total moment of 16 $\mu_B$. This value is 
close to the experimentally measured value of 1.72 $\pm$ 0.04 $\mu_B$/atom\cite{mark1, mark2}.
Recently Parvanova {\it et al.} predicted the same geometrical structure but with 
comparatively smaller, 1 $\mu_B$/atom, magnetic moment\cite{bobadova2}. We have found two closely lying
isomers, which have the same geometrical structure with total moments 4 $\mu_B$ and 
18 $\mu_B$ (Table \ref{tab:bemag}). Another possible icosahedral structure 
without the central atom lies much higher in energy.

The obvious candidates for the Mn$_{13}$ cluster are the icosahedral, hexagonal close packed (HCP)
and cuboctahedral structures. The variation of total energy as a function of the total 
magnetic moment is plotted in the Fig.\ref{fig:mn13} for all these three conformations.
The icosahedral structure is found to be the ground state 
with $N_{\uparrow}$ = 7 spin structure. The two pentagonal rings are AFM
coupled for this structure (Fig.\ref{fig:intermediate}).  
Consequently, the magnetic moment is found to be small, 
0.23 $\mu_B$/atom. This predicted magnetic moment is much smaller than those of its 
neighboring Mn$_{12}$ and Mn$_{14}$ clusters (Fig.\ref{fig:magmom}), what has been indeed predicted by the 
SG experiment\cite{mark1, mark2}. Although, the present value is much lower than the 
experimental value of 0.54 $\pm$ 0.06 $\mu_B$/atom\cite{mark2}. However, we have found another
icosahedral isomer with magnetic moment exactly the same with the experimental value, which  
lies only 0.08 eV higher in energy (Table \ref{tab:bemag}). This structure also has 
$N_{\uparrow}$ = 7. 
Recently, Parvanova {\it et al.} predicted similar magnetic ordering\cite{bobadova2}.
The optimal HCP and cuboctahedral structures (Fig.\ref{fig:intermediate})
 have relatively 
higher magnetic moments 9 $\mu_B$ ($N_{\uparrow}$ = 7) and 11 $\mu_B$ ($N_{\uparrow}$ = 8)
, respectively, and they lie much higher in energy, 0.89 eV and 1.12 eV, respectively.
Nayak {\it et al.} first predicted a ferrimagnetic state for Mn$_{13}$. However, their
predicted magnetic moment is quite high (33 $\mu_B$): all the surface atoms are 
antiferromagnetically coupled with the central atom\cite{nayak2}. 

The ground state of Mn$_{14}$ is the first complete icosahedra with a single atom capping.
This structure has $N_{\uparrow}$ = 9, with a magnetic moment 1.29 $\mu_B$/atom. In this
structure the magnetic coupling between the two pentagonal rings is FM, which 
was coupled antiferromagnetically in the case of Mn$_{13}$ and consequently, it has 
small magnetic moment. The next two isomers lie very close to the ground state:
they lie only 0.02 eV and 0.05 eV higher and have 1.43 and 1.57 $\mu_B$/atom magnetic
moment, respectively. These two isomers (not shown in Fig.\ref{fig:intermediate}) 
have the same $N_{\uparrow}$ = 9
spin structure, but with their different positional arrangement. 
The experimentally predicted magnetic moment, 
1.48 $\pm$ 0.03 $\mu_B$/atom, is an average of the ground state and the two isomers
(Table \ref{tab:bemag}), which again indicates that these isomers might be produced
along with the ground state in the SG experiment.       

The discrepancy between the present theoretical and experimental magnetic moment is rather large for
Mn$_{15}$. The present value is 0.87 $\mu_B$/atom, whereas the corresponding experimental
value is 1.66 $\pm$ 0.02 $\mu_B$/atom. We have also found several isomers (Table \ref{tab:bemag}), 
but none of them are close to the experimental value. The ground state and all the 
closely lying isomers within $\sim$ 0.1 eV energy spacing are of derived icosahedral
structure. The two competing icosahedral structures with 5,1,5,1,3 and 1,5,1,5,1,2 staking
(i.e. without or with the apex atom) are possible (Fig.\ref{fig:intermediate}). 
The first kind of structure is found to be the
ground state, whereas the optimal structure for the second kind lies 0.06 eV higher with a
magnetic moment 13 $\mu_B$ (Fig.\ref{fig:intermediate}). Another structure of the second kind
is found to be degenerate with this isomer, which has a magnetic moment 7 $\mu_B$ (Table \ref{tab:bemag}).
However, using US-PP Briere {\it et al.} found a bcc structure to be the 
ground state with much smaller magnetic moment, 0.20 $\mu_B$/atom\cite{briere}. In the present
case this bcc kind of structure lies 0.28 eV higher (Table \ref{tab:bemag}).  

The same structural trend is observed in the case of Mn$_{16}$, the two different competing geometries 
have been 
found to be the possible candidates for the Mn$_{16}$
cluster. Both of these structures can be derived from the 19-atom double icosahedra, which has
a 1,5,1,5,1,5,1-atomic staking. The ground state has a magnetic moment 1.25 $\mu_B$/atom with 
$N_{\uparrow}$ = 9 spin structure. This structure has 5,1,5,1,4-atomic staking: both the apex
atoms and one atom from the lower pentagonal ring are missing from the double icosahedra. The
next isomer has the same atomic arrangement and is almost degenerate, which lies only 0.02 eV higher. 
This has 1.38 $\mu_B$/atom
magnetic moment and the same ($N_{\uparrow}$ = 9) spin ordering. For both of these structures
the central pentagonal ring is antiferromagnetically coupled with the upper and lower (incomplete)
pentagonal ring.
The experimentally predicted value,    
1.58 $\pm$ 0.02 $\mu_B$,\cite{mark1, mark2} is very close to these predicted values and 
confirms the corresponding ground state to be really of this `strange' staking. The next two 
isomers have a different icosahedral geometry and have comparatively smaller magnetic moment, 0.63 
(Fig. \ref{fig:intermediate}) and 
0.50 $\mu_B$/atom. They lie 0.06 and 0.1 eV higher, respectively. Both of them have 
1,5,1,5,1,3 staking, i.e. the 13-atom icosahedra is complete. The two complete pentagonal
rings are antiferromagnetically coupled. All these structures have same number of 
$N_{\uparrow}$ and $N_{\downarrow}$ but have two different class of atomic arrangements, which 
is consequently the reason for their large difference in the magnetic moment.

The Mn$_{17}$ cluster follows the same structural trend seen in both Mn$_{15}$ and Mn$_{16}$. 
The ground state is a 
double icosahedra without both the apex atoms, i.e. it has 5,1,5,1,5 staking. The spin 
structure is $N_{\uparrow}$ = 10 and the central pentagonal ring is AFM coupled with the 
other rings. This structure has a magnetic moment of 1.59 $\mu_B$/atom, which is in excellent
agreement with the experiment, 1.44 $\pm$ 0.02 $\mu_B$/atom\cite{mark2}. The next two isomers
also have the same conformation as well as the same spin structure.
For this size the structure of the second kind i.e. the icosahedral structure with one apex
atom (Fig.\ref{fig:intermediate}) lies rather higher in energy.
To our knowledge, there is no available report for any other elements where this
kind of staking has been observed to be the ground state for Mn$_{15}$ $-$ Mn$_{17}$ clusters.

The Mn$_{18}$ is the 19-atom double icosahedra without one apex atom. The predicted magnetic 
moment is 1.67 $\mu_B$/atom, whereas the experimental value is slightly smaller, 1.20 $\pm$ 0.02
$\mu_B$/atom\cite{mark2}. Next two isomers are nearly degenerate and have 1.56 and 1.44 
$\mu_B$/atom magnetic moment (Table \ref{tab:bemag}). For all these structures the integrated 
magnetization densities
$\mathcal{M}$ for the central pentagonal bi-pyramid are negative (Fig. \ref{fig:intermediate}).

The double icosahedral conformation is found to be the ground state for Mn$_{19}$. The predicted 
magnetic moment is 1.10 $\mu_B$/atom (Fig. \ref{fig:intermediate}), which is 
smaller than those of its neighboring 
clusters, what has been predicted by the experiment\cite{mark1, mark2}. Another magnetic structure has been
found to be degenerate with 1 $\mu_B$/atom magnetic moment (not shown in Fig. \ref{fig:intermediate}). 
Both of the structures have $N_{\downarrow}$
= 7 and the central pentagonal ring is AFM coupled with the other two rings. However, the predicted 
magnetic moment is larger than the experimentally measured value, 0.41 $\pm$ 0.04 $\mu_B$/atom
\cite{mark2}. In our case a magnetic structure with a magnetic moment 0.47 $\mu_B$/atom  
($N_{\downarrow}$ = 9), which is very close to the experimentally measured value, lies only 
0.08 eV higher in energy (Fig \ref{fig:intermediate}). The optimal FCC structure lies much higher, 
1.53 eV, in energy, which is shown in Fig. \ref{fig:intermediate}.

Two degenerate ground states have been found with 1.40 and 1.50 $\mu_B$/atom magnetic moment
for Mn$_{20}$ cluster. Both the structures have $N_{\downarrow}$ = 7 (Fig. \ref{fig:intermediate}) 
spin configuration and the conformation can be seen as a singly capped 19-atom double icosahedra. 
The central pentagonal ring is antiferromagnetically coupled with the other two rings. 
The predicted ground state magnetic moment is larger than the
experimental value, 0.93 $\pm$ 0.03 $\mu_B$/atom\cite{mark2}. However, a different spin
structure ($N_{\downarrow}$ = 8) with magnetic moment 0.80 $\mu_B$/atom, which is close to the 
experimentally predicted value, lies only 0.07 eV higher
(Fig. \ref{fig:intermediate}).

In the intermediate size range, the grouping of like spin atoms i.e spin segregation 
occurs (Fig. \ref{fig:intermediate}). For a particular sized cluster, 
we find that the ferromagnetically
aligned atoms have longer average bond lengths \cite{average} than those of the antiferromagnetically 
aligned ones. This is because of the Pauli repulsion.

\subsection{\label{BE}Binding Energies}
\begin{figure}[!t]
{\rotatebox{270}{\resizebox{6cm}{8cm}{\includegraphics{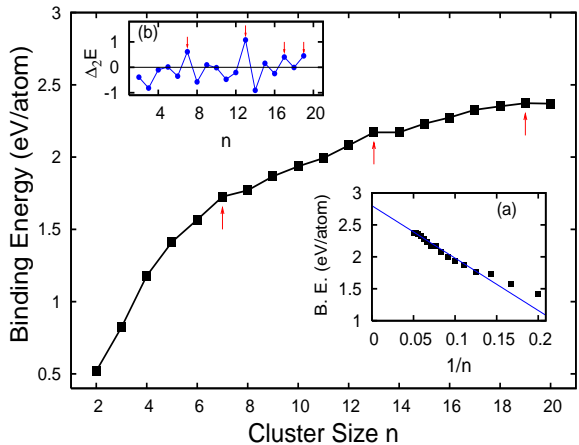}}}}
\caption{\label{fig:binding} (Color online) Plot of binding energy per atom as a
function of cluster size $n$ for the entire size range 2 $\le n \le$
20. (a) Plot of the same as a function of $1/n$ for the ferrimagnetic
clusters, 5 $\le n \le$ 20 and a linear fit (B.E. = $-$8.20$\frac{1}{n}$
+ 2.80) to the data.  (b) Plot of second difference, $\Delta_2E$ in
energy, which represents the relative stability.}
\end{figure}

The size dependence of the ground state
binding energy for Mn$_n$ clusters ($n=$2-20) is shown in Fig. \ref{fig:binding}.  Due to the lack of
hybridization between the half-filled 3$d$ and filled 4$s$ states and
due to high 4$s^2$3$d^5$ $\rightarrow$ 4$s^1$3$d^6$ promotion energy, the
Mn$_2$ dimer is a weakly bound dimer, which is a characteristic of van
der Waals bonding\cite{morse, gingerich, kant}. As the number of atoms in the cluster increases,
the binding energy increases monotonically due to the increase in the
$s-d$ hybridization. However, it remains weak as compared to the other
transition metal clusters in the same size range. This weak bonding has been demonstrated 
through the photodissociation experiments for Mn$_n^+$ ($n$ $\le$ 7) cluster 
cations\cite{teraski, tono}. Recently, we have shown \cite{kabir1} that if an As-atom 
is doped to the Mn$_n$ clusters, the binding energies of the resultant Mn$_n$As clusters
increase substantially due to their hybridized $s-d$ electrons bond
with the $p$ electrons of As. Similar enhancement in bonding has also been seen due to 
the single nitrogen doping\cite{jenaprl}. 

Upon extrapolation of the linear fit to the binding energy per atom
data to $1/n \rightarrow 0$ (Fig.\ref{fig:binding}(a)), we obtained
the binding energy for an infinitely large cluster as 2.80 eV, which
is very close to the experimental AF bulk $\alpha$-Mn (2.92 eV).  It
is important here to note the kinks observed at $n=$ 7 and 13 in the 
binding energy curve (Fig.\ref{fig:binding}). 
These kinks represent enhanced stability rendered by their   
`closed' geometric structures: Mn$_7$ is PBP and Mn$_{13}$ is the first complete
icosahedra. If this argument is valid then there should also be a kink
at $n=$ 19, due to the fact it has double icosahedral structure. But
we do not see any prominent kink in the binding energy curve. So, it will be interesting to
investigate the second difference in the binding energy, $\Delta_2 E(n)
= E(n+1)+E(n-1)-2E(n)$, where $E(n)$ represents the total energy of an
$n-$atom cluster. As $\Delta_2 E(n)$ represents stability of the
corresponding cluster compared to its neighbors, the effect will be
prominent. $\Delta_2E$ is plotted in the Fig.\ref{fig:binding}(b),
where we see a peak for Mn$_{19}$ too along with $n=$ 7 and 13.
However, in addition, without any a priory reason, we observe another
peak at $n=$ 17, which does not have `closed' structure (a double
icosahedra without two apex atoms).

\subsection{\label{trans}Transition in magnetic ordering}

\begin{figure}[!t]
{\rotatebox{270}{\resizebox{6cm}{8.5cm}{\includegraphics{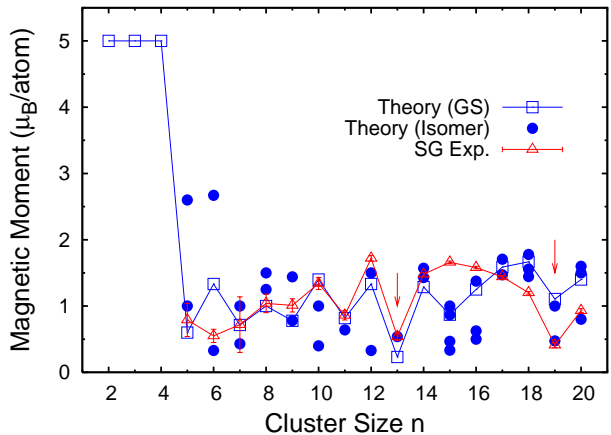}}}}
\caption{\label{fig:magmom} (Color online) Size dependent variation of magnetic moment. 
For the size range 5 $\le x \le$ 20, it shows excellent agreement with 
the SG experiment. Isomers which lie very close to the corresponding GS 
energy are also shown.}
\end{figure}

\begin{figure}[!b]
{\rotatebox{270}{\resizebox{6cm}{8.5cm}{\includegraphics{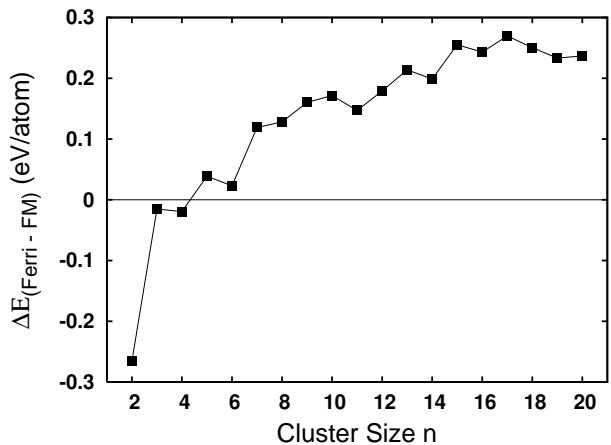}}}}
\caption{\label{fig:deltaE} Plot of $\triangle E_{(\rm{Ferri - FM})}$ as a function of cluster size $n$.}
\end{figure}

\begin{figure*}[!t]
{\rotatebox{0}{\resizebox{17cm}{7.5cm}{\includegraphics{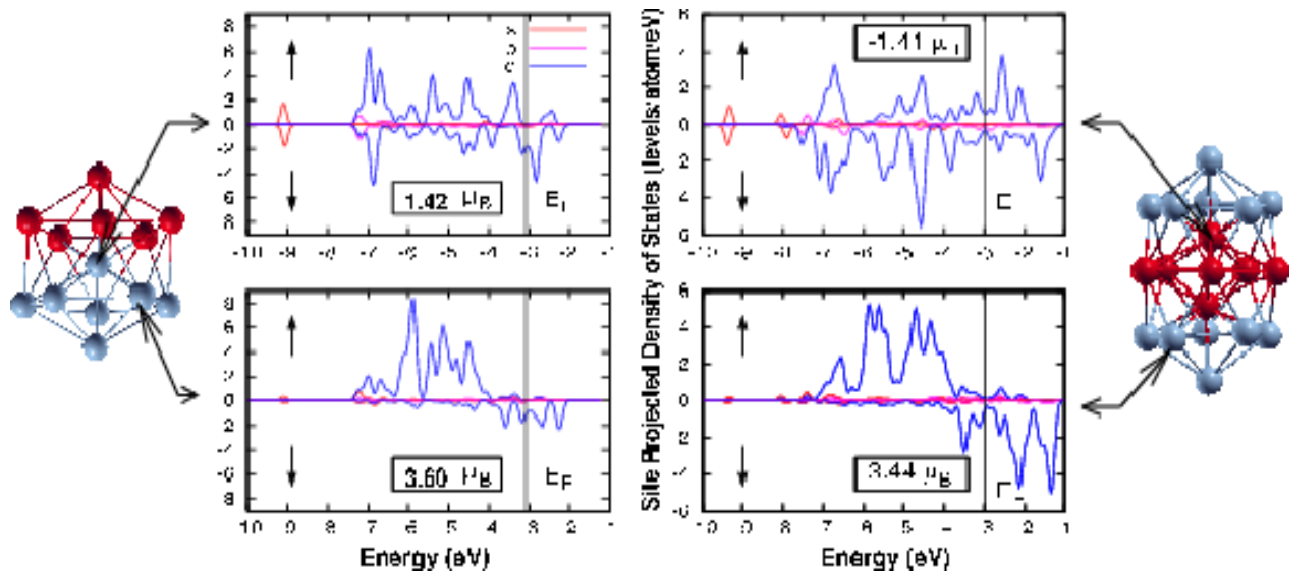}}}}
\caption{\label{fig:localization} (Color online) The $s$-, $p$- and $d$-projected density of 
states for the central and surface atoms for Mn$_{13}$ and Mn$_{19}$ in their ground state. Gaussian
broadening of half-width 0.1 eV has been used. Integrated magnetization density $\mathcal{M}$ 
for each atom is given in the box.}
\end{figure*}

For very small clusters, $n\le$4, the magnetic coupling is found to be FM with
magnetic moments 5 $\mu_B$/atom, which is the Hund's rule value for an
isolated Mn atom. Although we see that for Mn$_3$ cluster the FM
solution is nearly degenerate with the frustrated AFM
solution. The size dependence of the magnetic moment per atom is
plotted in the Fig.\ref{fig:magmom}. We see the transition in
the magnetic coupling (from FM to ferrimagnetic) takes place at $n=5$
and the ferrimagnetic states continue to be the ground state for
the entire size range $n=$5-20. 
Fig.\ref{fig:magmom} shows a very good agreement between experimentally
measured and our predicted magnetic moments. It was seen in the SG
experiment that the experimental uncertainty in measuring the magnetic
moment decreases with the cluster size. However, this is not the case
for Mn$_7$, for which the measured uncertainty is quite large (0.72$\pm$0.42
$\mu_B$, $\pm$58\% of the measured value)  as compared to the neighboing sizes. This
large uncertainty might arise from the presence of isomers with
different magnetic moments in the SG beam for subsequent measurements. 
However, in addition with the statistical fluctuation, the above explanation only 
stands for a plausible reason as for all other 
sizes we did find many isomers with different magnetic moments
(see Table \ref{tab:bemag} and Fig.\ref{fig:magmom}), but the
corresponding experimental uncertainty is not that large. 
One another striking feature observed in the experiment is the sudden
decrease in the magnetic moment at $n$=13 and 19, compared to their
neighbors. Our calculation reproduces this feature. This is attributed 
to their `closed' icosahedral structures: first complete icosahedra 
for Mn$_{13}$ and a double icosahedra for Mn$_{19}$. The other geometries studied
viz. hexagonal closed packed and cuboctahedral structures for Mn$_{13}$
and a fcc structure for Mn$_{19}$ lie much higher in energy, 
0.89 eV (9 $\mu_B$), 1.12 eV (11 $\mu_B$) and 1.53 eV (17 $\mu_B$), 
respectively, than their corresponding ground state.    

In the Fig.\ref{fig:magmom} we have depicted the magnetic moments of the very
closely lying isomers with their ground state (see Table \ref{tab:bemag}) and while comparing
those with the experimentally observed values, we come to the
conclusion that for a particular size of cluster, the isomers with
different magnetic moments are likely to be present in the SG cluster
beam with a statistical weight and essentially, the measured moment is
the weighted average of those isomers.
We calculate the 
energy difference between the optimal FM and optimal ferrimagnetic solutions,
$\triangle E_{(\rm{Ferri - FM})} = E({\rm{Ferri}}) - E({\rm{FM}})$,  and plot them as a function 
of cluster size $n$ in the Fig.\ref{fig:deltaE}. For both Mn$_3$ and Mn$_4$ the FM solutions are slightly lower
in energy than those of their respective optimal ferrimagnetic solutions, 
whereas the optimal FM solutions are slightly higher
than the corresponding ferrimagnetic ground states for Mn$_5$ and Mn$_6$. Thereafter, as the cluster size 
increases, this energy difference,  $\triangle E_{(\rm{Ferri - FM})}$, increases almost monotonically indicating 
that the optimal FM solutions become more and more unlikely. All these optimal FM states 
have $\sim$ 4 $\mu_B$/atom magnetic moments.

\subsection{\label{localization}Coordination and the $d$-electron localization}
The angular momentum projected local density of states (LDOS) show interesting site dependency. The $s$-, $p$-
and $d$-projected LDOS for the central and surface atoms are plotted in the Fig.\ref{fig:localization} for 
the Mn$_{13}$ and Mn$_{19}$ clusters. We see only $d$-projected LDOS are significant and are of great 
interest here. The $d$-projected LDOS of both Mn$_{13}$ and Mn$_{19}$ for the central atoms are broad for
both majority and minority spin states, which are also reflected through their small values of the integrated 
spin densities $\mathcal{M}$ (1.42 and -1.41 $\mu_B$ for Mn$_{13}$ and Mn$_{19}$, respectively). The broadening occurs due
to the high coordination of the central atom. On the other hand, the $d$-projected LDOS of the surface atoms
are rather localized and the majority spins are nearly fully occupied, which is in agreement with the 
relatively large local magnetic moments of the surface atoms (3.60 and 3.44 $\mu_B$ for Mn$_{13}$ and Mn$_{19}$
, respectively).

\subsection{\label{spingap}Spin gaps: Nonmetal $-$ metal transition?}
\begin{figure}[!t]
{\rotatebox{270}{\resizebox{6cm}{8.5cm}{\includegraphics{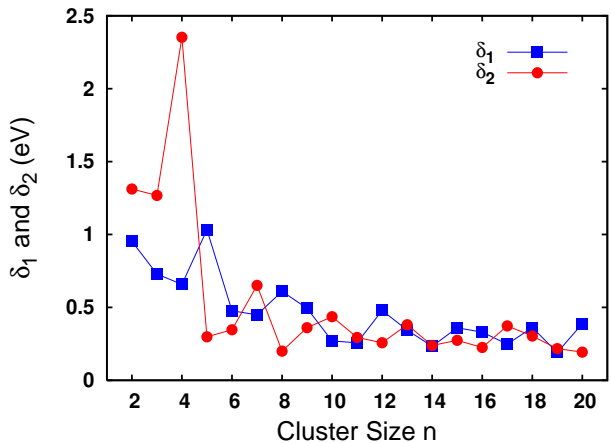}}}}
\caption{\label{fig:spingaps} (Color online) Plot of spin gaps as a function of cluster size $n$. 
See Table \ref{tab:bemag} for the numerical values.}
\end{figure}

A spin arrangement in any magnetic clusters is magnetically stable only if both the spin gaps, 
\begin{eqnarray}
\delta_1 \ &=& \ - \ \left[ \ \epsilon_{\rm HOMO}^{\rm majority} \ - \ 
                                \epsilon_{\rm LUMO}^{\rm minority} \right]
\nonumber\\
\delta_2 \ &=& \ - \ \left[ \ \epsilon_{\rm HOMO}^{\rm minority} \ - \ 
                               \epsilon_{\rm LUMO}^{\rm majority} \right],
\end{eqnarray}
are positive, i.e. the lowest unoccupied molecular orbital (LUMO) of the majority spin lies above 
the highest occupied molecular orbital (HOMO) of the minority spin and vice versa. We find these two
spin gaps to be positive for all the clusters (Table \ref{tab:bemag}) discussed here and are plotted
in the Fig.\ref{fig:spingaps}. 
$\delta_1$ and $\delta_2$ have local structures, but generally decreases slowly as the coordination increases with cluster size. 
Parks {\it et al.} found that the Mn$_n$ clusters with $n \le$ 15 are not reactive  towards molecular hydrogen, 
whereas they form stable hydrides at and above $n =$ 16, and the reaction rate varies considerably with the cluster size\cite{parks}.
 They argued it to be attributed from the 
non-metal to metal transition at $n =$ 16. If this is indeed the reason, it is likely that the 
ionization potential would show a significant decrease at Mn$_{16}$, similar to what has been 
observed for free mercury clusters\cite{rademann}. Therefore, we expect closing up of the spin gaps at $n =$ 16.
However, Koretsky {\it et al.} observed no sudden decrease in the measured ionization 
potential \cite{koretsky} and we 
do not find any spin gap closing at Mn$_{16}$ either. The spin gaps have reasonable value, 
$\delta_1 =$  0.33 eV and $\delta_2 =$ 0.22 eV for Mn$_{16}$ cluster (Fig.\ref{fig:spingaps} and Table 
\ref{tab:bemag}). 
This abrupt change in the 
reaction rate with H$_2$ at Mn$_{16}$ is not due to any structural change either, as we find all the 
medium sized clusters adopt icosahedral growth pattern and the reason for the observed change in the 
reaction rate remains unknown.    

\section{\label{summary}Summary and Conclusions}

We have systematically investigated the structural, electronic and magnetic
properties of Mn$_n$ ($n$ = 2-20) clusters from the first-principles density
functional theory. An extensive search have been made to locate the global
minima. Due to the intrinsic 4$s^2$
3$d^5$ electronic structure and high 4$s^2$3$d^5$  $\rightarrow$ 4$s^1$3$d^6$
promotion energy Mn-atoms do not bind strongly when they come closer to form
a cluster. However, binding energy increases with the cluster size as the
coordination number increases and reaches a value 2.37 eV/atom for Mn$_{20}$,
which is 81 \%
of the bulk value. A magnetic transition from FM to ferrimagnetic ordering  
takes place
at $n$ = 5 and thereafter the energy difference between the optimal ferrimagnetic and
optimal FM
structure increases with the cluster size, which indicates that the optimal 
FM states become more and more unfavorable with increasing cluster size . 
However, different ferrimagnetic states
are possible
within a small $\sim$ 0.1 eV energy difference and their plausible presence
in the experimental SG beam along with the ground state has been argued. The
predicted magnetic moments are in  agreement with the SG experiment.
The sudden decrease in the
magnetic moment at $n$ = 13 and 19 is due to their `closed' icosahedral
structure. It should be pointed out here that in the present calculation
we assumed only collinear alignment of spins. However, spin canting or
noncollinear magnetic ordering is possible in small magnetic clusters as
it occurs more easily in a low symmetry magnetic system\cite{kabir1}. 
Icosahedral growth pattern is observed for the intermediate size
range. However, to our knowledge, a different kind of icosahedral packing 
have been observed for Mn$_{15}$ $-$ Mn$_{17}$ clusters. 
In any particular cluster, the average bond length between antiferromagnetically 
aligned atoms are 3$-$8 \% shorter than that of the ferromagnetically alligned, 
which can be explained in terms of the Pauli repulsion. Spin segregation has been observed 
in the intermediate size range. 
The $d-$electron localization strongly depends on 
coordination: localization decreases with the coordination number.
There is no signature of non-metal to metal transition at 
$n$ =16, which has been predicted\cite{parks} through the downward discontinuity observed in the 
reaction rate with H$_2$.

\acknowledgments
We are grateful to Dr. P. A. Sreeram for helpful discussions. 
This work has been done under the Indian Department of Science and Technology 
grant No. SR/S2/CMP-25/2003.
M. K. thankfully acknowledges the congenial hospitality at the Centre for Modeling and Simulation of Pune
University. We also acknowledge the DST grant No. SP/S2/M-20/2001 for using computational resource.


\begin{thebibliography}{99}
\bibitem{average}The average bond lengths were calculated by assuming a cut-off of 3 \AA.
\end{thebibliography}
\end{document}